\documentclass[a4paper,11pt]{article}
\usepackage{pos}
\usepackage{amsmath}

\newcommand{\overleftrightsmallarrow}{\mathpalette{\overarrowsmall@\leftrightarrowfill@}}

\title{Standard Model and SMEFT Higgs theory overview}

\author*[a]{Ramona Gr\"ober}

\affiliation[a]{Universit\`a di Padova and INFN Sezione di Padova,\\
  Via F Marzolo, Padova, Italy}

\emailAdd{ramona.groeber@pd.infn.it}

\abstract{After more than a decade from its discovery, the Higgs boson remains at the centre of the particle physics programme. 
While its couplings to vector bosons and third-generation fermions have been measured with impressive precision, 
the structure of the Higgs potential, the self-couplings, and the interactions with first- and second-generation fermions 
are still poorly constrained. 
This contribution summarises recent theoretical progress in precision predictions in Higgs physics, 
the interpretation of these results in Effective Field Theories (EFTs), 
and the current and future prospects to probe light Yukawa couplings at the LHC and beyond.}

\FullConference{The European Physical Society Conference on High Energy Physics (EPS-HEP2025)\\
7-11 July 2025\\
Marseille, France\\}

\begin{document}
\maketitle

\section{Introduction}
The situation in particle physics is, in many respects, more puzzling than ever.  With the discovery of the Higgs boson in 2012~\cite{ATLAS:2012yve,CMS:2012qbp}, the last missing ingredient of the Standard Model (SM) was finally observed, 
yet we still lack clear guidance on how to resolve the deep puzzles the SM leaves unanswered.  
Since the Higgs boson is directly connected to several of these open questions --
such as the hierarchy problem, the stability of the electroweak vacuum, 
and possibly also the origin of the matter-antimatter asymmetry of the Universe -- 
ensures that Higgs physics remains a vibrant and exciting field.  
A wealth of precision measurements of Higgs-boson properties 
will become available in upcoming runs of the LHC and its high-luminosity upgrade (HL-LHC)~\cite{Cepeda:2019klc}.

The Higgs couplings to massive gauge bosons and third-generation fermions 
have already been measured with remarkable precision~\cite{ATLAS:2022vkf,CMS:2022dwd}.  
In contrast, the Higgs self-couplings are still only constrained to be within several times their SM value \cite{ATLAS:2025hhd, CMS:2025ngq}, 
and the Yukawa couplings to first- and second-generation fermions remain largely unknown, 
with the notable exception of the muon coupling~\cite{CMS:2020xwi,ATLAS:2025coj}.  
More generally, the effective Lagrangian describing potential deviations from the SM 
still allows for a broad range of possibilities that need to be systematically explored.

A comprehensive strategy to investigate the Higgs sector is build on two pillars.  
First, precise higher-order predictions within the SM are crucial to provide accurate and reliable benchmarks for data comparisons.  
Second, one should explore in a systematic and theoretically consistent way 
which kinds of deviations from the SM are possible and how they can be probed experimentally.  
In the absence of direct signals of new physics, 
Effective Field Theories (EFTs) such as the Standard Model EFT 
(SMEFT)~\cite{Buchmuller:1985jz,Grzadkowski:2010es} 
and the Higgs EFT (HEFT)~\cite{Feruglio:1992wf,Alonso:2012px,Buchalla:2013rka} 
provide a powerful and model-independent framework to describe heavy new states.  
Nevertheless, EFT interpretations should be complemented by checks of their 
possible ultraviolet (UV) realisations to ensure that the inferred deviations 
are consistent with well-defined extensions of the SM~(see e.g. Refs.~\cite{DiLuzio:2017tfn, Durieux:2022hbu} for trilinear Higgs self-coupling deviations).

This contribution discusses, in Section~\ref{sec:precision}, 
the current status of precision Higgs physics; 
in Section~\ref{sec:EFT}, the application of SMEFT and HEFT 
to the interpretation of Higgs observables; 
and, in Section~\ref{sec:Yuk}, concrete model realisations 
that can induce deviations in the light Yukawa couplings.

\section{Precision predictions in Higgs physics \label{sec:precision}}
Precise theoretical predictions for Higgs physics might become the bottleneck for the HL-LHC program. In projections usually a halving of the theory uncertainty is already assumed, nevertheless in various Higgs production and decay channels the theory uncertainty might turn out to be the dominant one at the HL-LHC \cite{ATLAS:2025eii}.

In the recent years, due to the development of new tools and methods, an impressive level of precision in theoretical predictions for Higgs production and decay have been reached.  

For the dominant gluon-fusion channel, QCD corrections are known up to N$^3$LO in the heavy-top limit \cite{Anastasiou:2016cez, Mistlberger:2018etf}, electroweak corrections at NLO \cite{Aglietti:2004nj, Actis:2008ug} and the mixed QCD+EW corrections up to $\mathcal{O}(\alpha_s^3 \alpha^2)$\cite{Anastasiou:2018adr}. One of the recent developments is the computation of
finite-quark-mass effects at NNLO and the interference of top and bottom loops that allowed also to study the renormalisation scheme dependence~\cite{Czakon:2023kqm, Czakon:2024ywb}.  
These results, together with the advent of approximate N$^3$LO PDF sets \cite{McGowan:2022nag, Cridge:2024icl, NNPDF:2024nan}
and refined missing-higher-order estimates, 
are crucial to control theoretical uncertainties in the HL-LHC era.

Gluon fusion is also relevant for other Higgs production processes, ranging from double Higgs production to the gluon-induced $Zh$ production to off-shell Higgs production with sizeable interference of the continuum $ZZ$ production in gluon fusion with Higgs production and subsequent decay to $ZZ$.
Due to the multiple scales in the problem, the calculation of higher order corrections is more difficult with respect to the single Higgs case. 
Taking the example of double Higgs production, which plays a central role in probing 
the Higgs self-coupling $\lambda_{hhh}$, various methods of the computation of two-loop multi-scale computations were tested. The computation of the full NLO QCD corrections with exact top-mass dependence 
has been performed numerically in~\cite{Borowka:2016ehy, Baglio:2018lrj}. Analytically, a possibile approach is the one of expansions, like the heavy-top mass expansion \cite{Dawson:1998py, Grigo:2013rya, Degrassi:2016vss} limited to the phase space $\hat{s}< 4 m_t^2$, where  $\hat{s}$ denoted the partonic center of mass energy and $m_t$ the top quark mass. The expansion can be improved in combination with a threshold expansion and Pad\'e approximants \cite{Grober:2017uho}. An expansion in small transverse momentum as proposed in \cite{Bonciani:2018omm} covers more than 95\% of the phase space. For the rest of the phase space, this expansion can be combined with a high-energy expansion \cite{Davies:2018qvx}, see \cite{Bellafronte:2022jmo, Davies:2023vmj} and hence can give a fully analytic description of the full phase space. The thus obtained matrix elements are very fast in their numerical evaluation and were included into a \textsc{Powheg} code \cite{Bagnaschi:2023rbx}. They are fully flexible, so allow to change the top mass renormalisation allowing to evaluate the corresponding uncertainty \cite{Baglio:2020wgt}. In particular, at large invariant masses the scheme dependence is especially pronounced and can be up to $\sim 40 \%$ for some bins in the invariant mass of the Higgs boson pair, $M_{HH}$. This uncertainty is currently the main player in the total uncertainty budget of Higgs pair production in gluon fusion. Recently, Ref.~\cite{Jaskiewicz:2024xkd} helped in the understanding of the origins of logarithms in the top mass in the high-energy limit via SCET. This could in the future be a street ahead in reducing the top mass renormalisation scheme uncertainty.  Alternatively, a reduced uncertainty will be obtained once mass dependent results at NNLO QCD are available. First efforts in this direction \cite{Davies:2023obx, Davies:2024znp, Davies:2025ghl} make use of the expansion in small transverse momentum. 
\begin{figure}
\centering
\includegraphics[width=10cm]{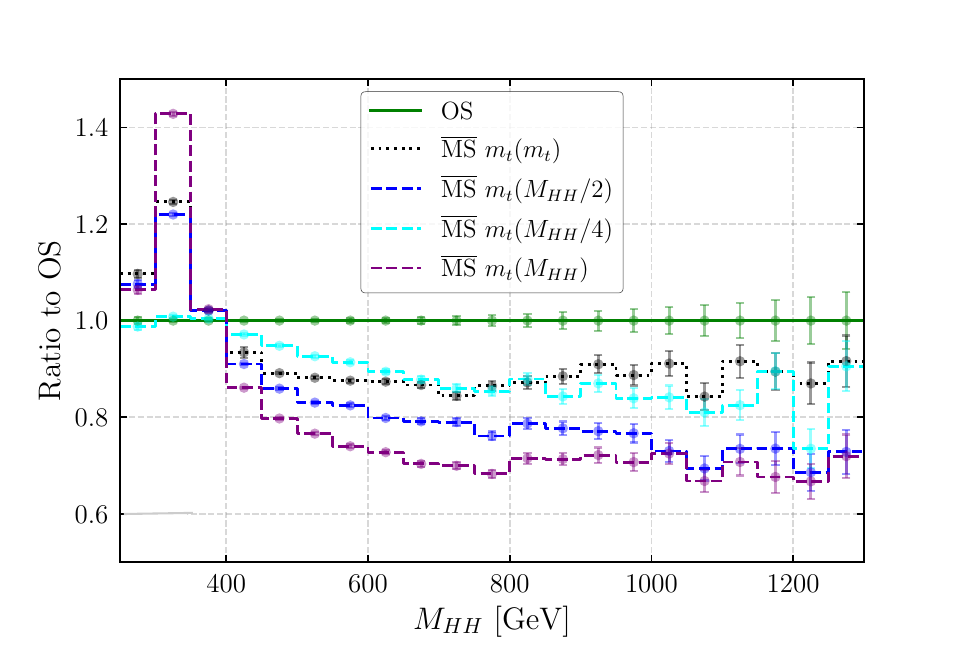}
\caption{Ratio of various choices of the top mass renormalisation scheme with respect to the onshell (OS) prediction. Figure taken from Ref.~\cite{Bagnaschi:2023rbx}.}
\end{figure}

\section{Effective Field Theories in Higgs Physics \label{sec:EFT}}

\begin{figure}
\includegraphics[width=7cm]{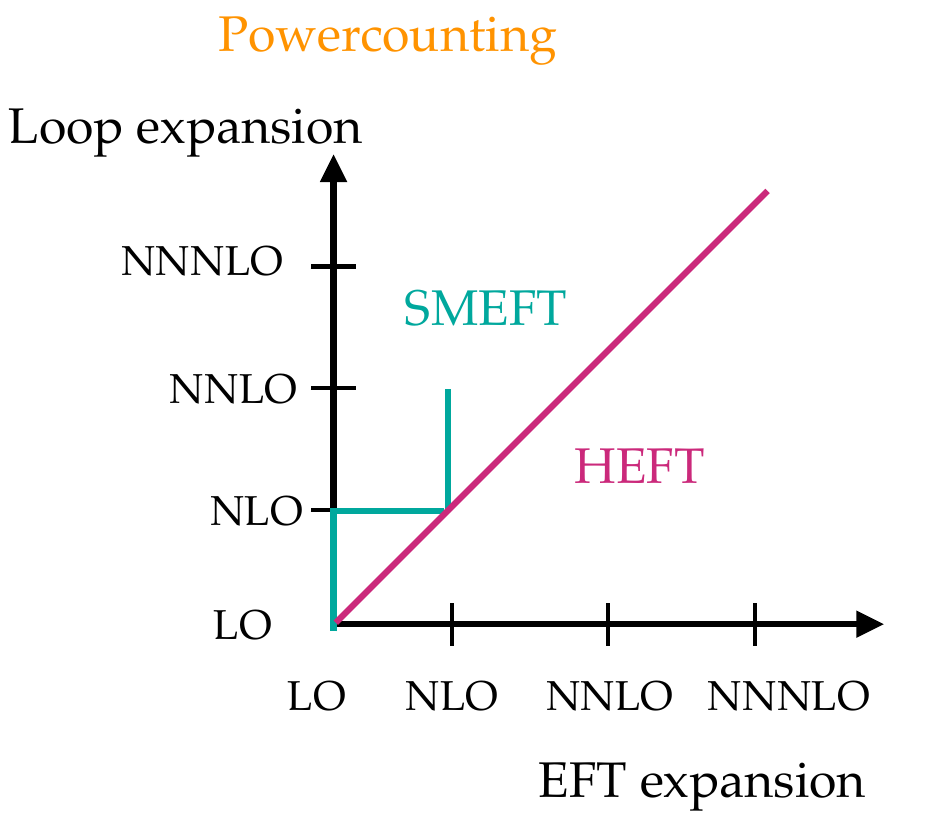}
\includegraphics[width=8cm]{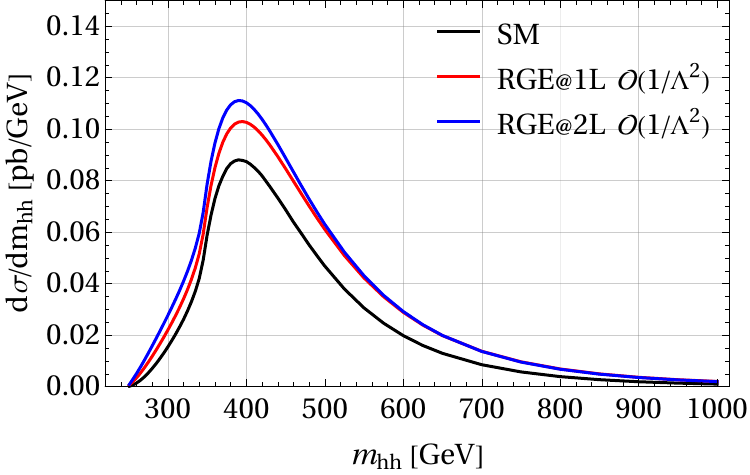}
\caption{\textit{Left:} Schematic description of the powercounting in SMEFT and HEFT. In SMEFT loop expansion and EFT expansion are independent, while in HEFT they are not, so in a consistent expansion one should stay on the diagonal. \textit{Right:} Example of the impact of the RGE running of the SMEFT Wilsone coefficients in Higgs pair production. Figure taken from Ref.~\cite{DiNoi:2024ajj}. \label{fig:mhhRGE}}
\end{figure}

Effective Field Theories (EFTs) provide a systematic description of possible deviations 
from the SM in a model-independent way. From the bottom-up perspective, the construction of an EFT describing new physics effects requires as input the particle content, the symmetries and a truncation (``power-counting'') rule a required. In the SMEFT, particle content and symmetries follow the SM and  
the Lagrangian is organised as
\begin{equation}
\mathcal{L}_\text{SMEFT} = \mathcal{L}_\text{SM} + 
\sum_i \frac{C_i}{\Lambda^2} \mathcal{O}_i \, ,
\end{equation}
where $\Lambda$ denotes the scale of new physics. 
In HEFT, the Higgs boson no longer comes together with the Goldstone bosons in a Higgs doublet but transforms instead as a singlet. The prize to pay is that the theory becomes non-unitary for scales $\Lambda> 4 \pi v$, where $v\approx 246\text{ GeV}$ denotes the SM vacuum expectation value.
As a consequence the power-counting rule cannot count orders of $\Lambda$ but is more complicated. Instead, the chiral dimension \cite{Buchalla:2013eza} can be counted, with Ref.~\cite{Brivioinprep} demonstrating on how to propagate the counting from Lagrangian to observable, intertwining loop and EFT expansion as shown schematically in Fig.~\ref{fig:mhhRGE} left.  
\par
In what regards the Higgs pair production process, HEFT and 
SMEFT operators at NLO QCD have been implemented in the  \textsc{Powheg} program \texttt{gghh}~\cite{Heinrich:2020ckp, Heinrich:2022idm}. 
The leading loop SMEFT effects can instead be estimated considering the effect of renormalisation group equation (RGE) running of the Wilson coefficients, see for the size of the effect in Higgs pair production~\cite{Maltoni:2024dpn, Heinrich:2024rtg}.
Since the operator $\mathcal{O}_{HG}=H^{\dagger}H G_{\mu\nu}G^{\mu\nu}$ contributes at leading order (LO) to Higgs pair production and the SM at one-loop level, an analysis of the dominant RGE effects should also include the  two-loop RGEs of $\mathcal{O}_{HG}=H^{\dagger}H G_{\mu\nu}G^{\mu\nu}$ from potentially tree-level generated operators for $\mathcal{O}_{HG}$  as computed in~\cite{DiNoi:2023ygk, DiNoi:2024ajj,DiNoi:2025tka}. An example of their relevance can be seen in Fig.~\ref{fig:mhhRGE}. It should also be noted that the two-loop RGE of $\mathcal{O}_{HG}$ depends on the continuation scheme of $\gamma_5$ in $d$ dimensions \cite{DiNoi:2023ygk}.\footnote{The $\gamma_5$ scheme dependence was also shown in the two-loop quark and gluon field renormalisation constants contributions of four-top operators \cite{DiNoi:2025arz, Duhr:2025yor}.} While most automated tools in SMEFT use the naive scheme \cite{Chanowitz:1979zu}, the Breitenlohner-Maison-t`Hooft-Veltman scheme (BMHV) \cite{tHooft:1972tcz, Breitenlohner:1977hr} is the only known algebraically consistent one. Recent works \cite{DiNoi:2025uan, Fuentes-Martin:2025meq} provide a first step towards translation between the two schemes. Bottom-up fits to data depend on the scheme choice of $\gamma_5$ \cite{DiNoi:2025uhu}. 
In conclusion, a consistent SMEFT interpretation must also address truncation uncertainties, 
renormalisation-scheme and $\gamma_5$ continuation effects, 
and the interplay between loop and EFT expansions.


The HEFT framework is in particularly interesting for multi-Higgs production as interactions with fermions, gluons or vector bosons and several Higgs bosons are decorrelated from the single Higgs couplings so HEFT seems to provide a more general prediction for Higgs pair production.\footnote{As shown in \cite{Grober:2025vse} this is a question of convergence of the EFT series: all structures in HEFT arise also in SMEFT but potentially at higher orders in the EFT expansion.} In Ref.~\cite{Brivioinprep2} the impact of HEFT NLO operators on Higgs pair production are investigated using a consistent power counting \cite{Brivioinprep}. The NLO HEFT operators can lead to new type of kinematic distributions so far not considered in the shape benchmarks \cite{Carvalho:2015ttv, Capozi:2019xsi, Alasfar:2023xpc} used by the experimental collaborations.
Given the large theory and experimental uncertainties, it is though rather unlikely to find kinematic distributions in the HEFT parameter space that have not yet been considered. Should the uncertainty decisively shrink with respect to current HL-LHC projections, the kinematic benchmarks might no longer be sufficient.

\section{Light generation Yukawa couplings \label{sec:Yuk}}
Despite the remarkable success in measuring third-generation Yukawa couplings, the Higgs interactions with first- and second-generation fermions remain largely unconstrained.  
Even future 
HL-LHC projections from global fits show that they can be constrained only much above their SM value
$|\kappa_c| < 1.2$, $|\kappa_s| < 13$, $|\kappa_d| < 260$, and $|\kappa_u| < 560$ at 95\% CL 
\cite{deBlas:2019rxi}, where $\kappa_q=g_{hq\bar{q}}/g_{hq\bar{q}}^{\text{SM}}$ describes the ratio to the SM coupling.  
Alternative proposals to constrain them exploit the Higgs transverse-momentum spectrum \cite{Bishara:2016jga,Soreq:2016rae}, 
the $W^\pm h$ charge asymmetry \cite{Yu:2016rvv}, or associated $h+\gamma$ \cite{Aguilar-Saavedra:2020rgo} production, triboson production \cite{Falkowski:2020abc}, Higgs pair production \cite{Alasfar:2019pmn, Alasfar:2022vqw} and off-shell Higgs production \cite{Balzani:2023jas}. The latter shows the best HL-LHC projections for the up- and down-coupling, $|\kappa_d|< 156$ and $|\kappa_u|< 260$. 

Given that the projections show that these couplings can only be constrained very weakly, a natural question to pose is whether there are any models that would predict such large deviations. A possible way to address this question is to study models that generate the SMEFT operator that leads to a modification of the light quark couplings to the Higgs boson at tree-level. 
The only model with a single mediator is the two-Higgs doublet model \cite{Egana-Ugrinovic:2019dqu}, but in case of two mediators the operator of type 
$\bar q_L H q_R (H^\dagger H)$ can be generated by models with two heavy vector-like states or one vector-like state and new scalars, or a second Higgs doublet and further scalar states \cite{Bar-Shalom:2018rjs, Nir:2024oor,Giannakopoulou:2024unn, Erdelyi:2024sls}. 
The models with two different representations of VLQs have the advantage that they avoid di-jet resonances. Apart from generating operators of type $H^3 \bar{\psi}\psi$ also operators of type $(H^{\dagger}\overset{\leftrightarrow}{D}H)(\bar{\psi} \psi)$ are generated at tree-level, where $\psi$ stands generically for a fermion. Furthermore, they generate various operators at one-loop level which can be constrained by electroweak precision tests, Higgs physics or flavour physics. Taking those constrains and future projections from the HL-LHC into account one can indeed find that tests like \cite{Balzani:2023jas} probe interesting model parameter space \cite{Erdelyi:2024sls}. Furthermore, one can find that these models can be tested well at the FCC-ee $Z$-pole run, which will leave much reduced range for light quark Yukawa enhancements. 

\begin{figure}
\centering
\includegraphics[width=7cm]{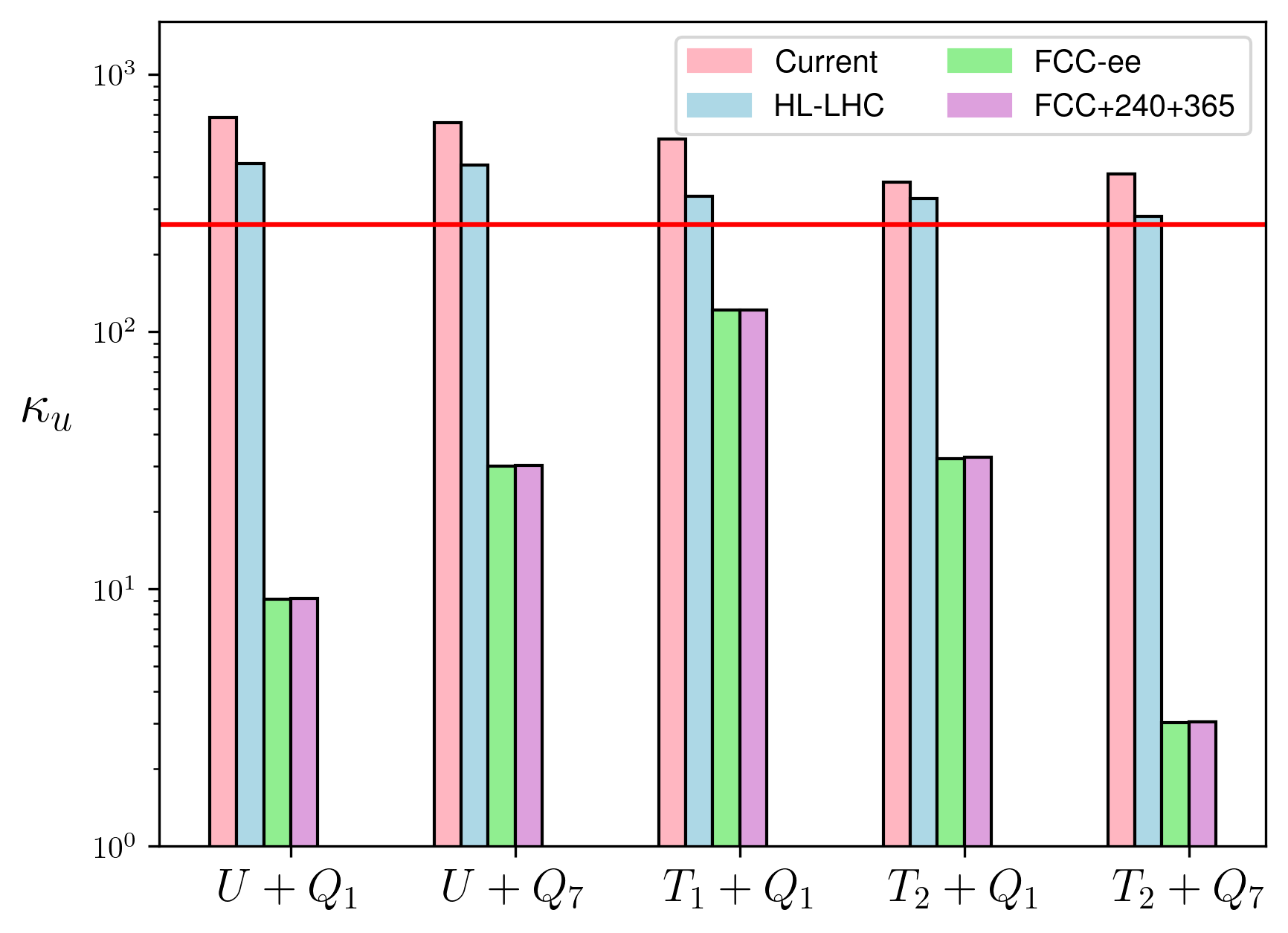}
\includegraphics[width=7cm]{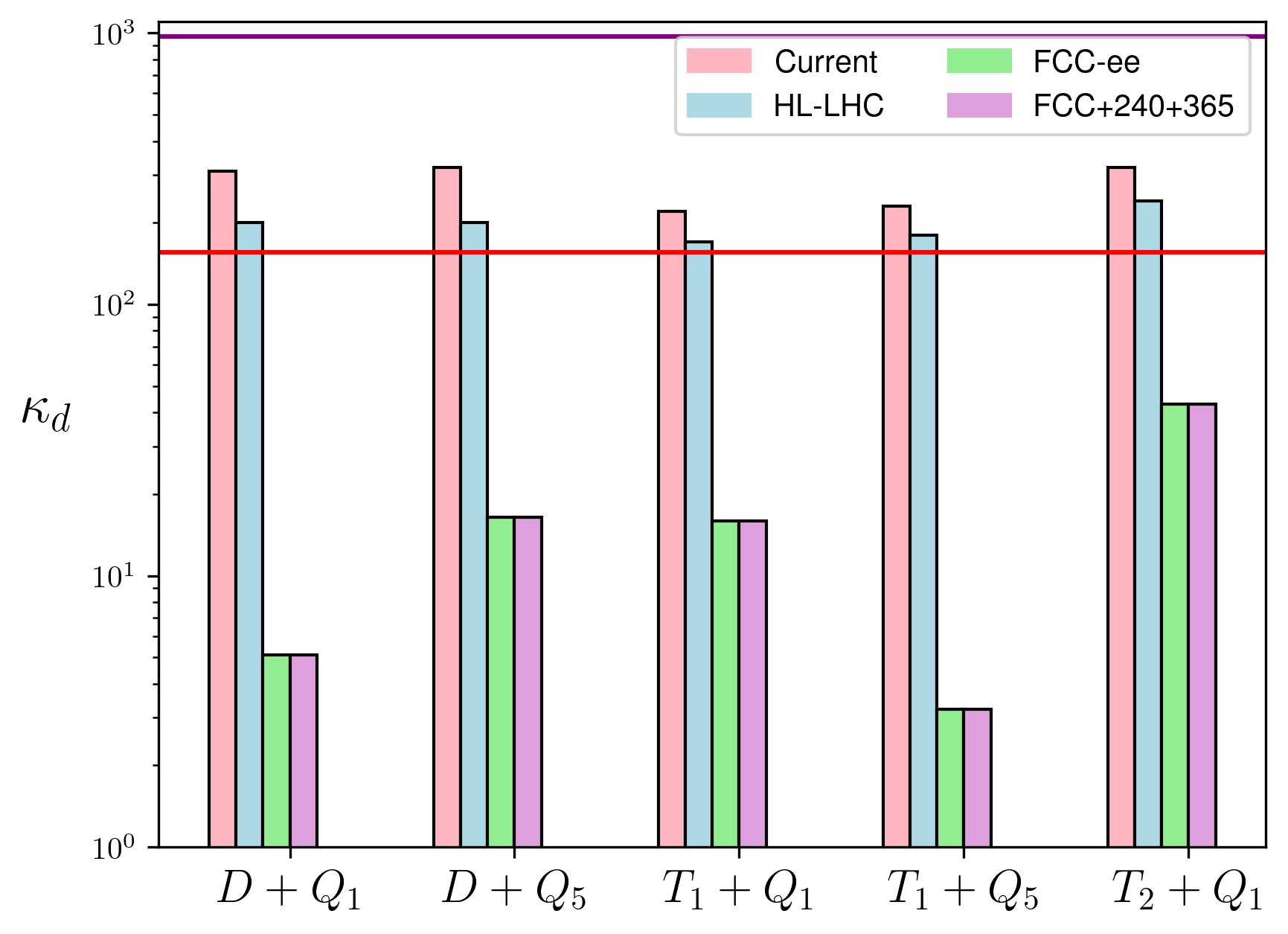}
\caption{Maximally allowed values of the light quark Yukawa couplings in various models featuring two vector-like quark states (the notation for those is following Ref.~\cite{deBlas:2017xtg}). The red line indicates HL-LHC projections from  ~\cite{Balzani:2023jas}, whereas the purple line are CMS bounds~\cite{CMS:2025xkn}.  Figure taken from Ref.~\cite{Erdelyi:2024sls}. \label{fig:kappaq}}
\end{figure}

Analogous constructions with vector-like leptons or scalars can enhance the electron Yukawa coupling \cite{Davoudiasl:2023huk, Erdelyi:2025axy, Allwicher:2025mmc},
potentially testable in a dedicated FCC-ee run at the Higgs pole 
with projected reach $\kappa_e < 1.6$~\cite{dEnterria:2021abc}.

\section{Conclusion}

Recent years have brought substantial progress in our understanding of the Higgs sector. On the theory side, ever more precise predictions have become essential and, given the absence of clear new-physics signals, EFTs have emerged as indispensable tools for Higgs characterisation. The SMEFT framework has seen remarkable developments, with the RGE of the Higgs gluon interaction now computed at (partial) two-loop precision. The more general HEFT formalism, although less explored, offers increased flexibility, especially for describing multi-Higgs dynamics where SMEFT may be overly restrictive. Efforts on consistent power-counting schemes aim to render HEFT more systematic and automatable. Finally, viable UV completions can accommodate enhanced first- and second-generation Yukawa couplings, providing a well-motivated target for the challenging searches in this sector.


\begin{thebibliography}{99}

\bibitem{ATLAS:2012yve}
ATLAS Collaboration, 
\emph{Phys.\ Lett.\ B} {\bf 716} (2012) 1 [arXiv:1207.7214].

\bibitem{CMS:2012qbp}
CMS Collaboration, 
\emph{Phys.\ Lett.\ B} {\bf 716} (2012) 30 [arXiv:1207.7235].

\bibitem{Cepeda:2019klc}
M.~Cepeda \textit{et al.},
\emph{CERN Yellow Rep.\ Monogr.} {\bf 7} (2019) 221 [arXiv:1902.00134].

\bibitem{ATLAS:2022vkf}
G.~Aad \textit{et al.} [ATLAS],
Nature \textbf{607} (2022) no.7917, 52-59
[erratum: Nature \textbf{612} (2022) no.7941, E24]
[arXiv:2207.00092 [hep-ex]].

\bibitem{CMS:2022dwd}
A.~Tumasyan \textit{et al.} [CMS],
Nature \textbf{607} (2022) no.7917, 60-68
[erratum: Nature \textbf{623} (2023) no.7985, E4]
[arXiv:2207.00043 [hep-ex]].



\bibitem{ATLAS:2025hhd}
G.~Aad \textit{et al.} [ATLAS],
[arXiv:2507.03495].

\bibitem{CMS:2025ngq}
 CMS Collaboration,
[arXiv:2510.07527].

\bibitem{ATLAS:2025coj}
G.~Aad \textit{et al.} [ATLAS],
[arXiv:2507.03595].

\bibitem{CMS:2020xwi}
CMS Collaboration, 
\emph{Phys.\ Rev.\ Lett.} {\bf 125} (2020) 061801.


\bibitem{Buchmuller:1985jz}
W.~Buchmüller and D.~Wyler, 
\emph{Nucl.\ Phys.\ B} {\bf 268} (1986) 621.


\bibitem{Grzadkowski:2010es}
B.~Grzadkowski, M.~Iskrzynski, M.~Misiak and J.~Rosiek, 
\emph{JHEP} {\bf 10} (2010) 085
[arXiv:1008.4884].

\bibitem{Feruglio:1992wf}
F.~Feruglio, 
\emph{Int.\ J.\ Mod.\ Phys.\ A} {\bf 8} (1993) 4937 [hep-ph/9301281].


\bibitem{Alonso:2012px}
R.~Alonso, M.~B.~Gavela, L.~Merlo, S.~Rigolin and J.~Yepes, 
\emph{Phys.\ Lett.\ B} {\bf 722} (2013) 330 [arXiv:1212.3305].

\bibitem{Buchalla:2013rka}
G.~Buchalla, O.~Cat{\`a} and C.~Krause,
Nucl. Phys. B \textbf{880} (2014), 552-573
[erratum: Nucl. Phys. B \textbf{913} (2016), 475-478]
[arXiv:1307.5017].


\bibitem{DiLuzio:2017tfn}
L.~Di Luzio, R.~Gr{\"o}ber and M.~Spannowsky,
Eur. Phys. J. C \textbf{77} (2017) no.11, 788
[arXiv:1704.02311 ].

\bibitem{Durieux:2022hbu}
G.~Durieux, M.~McCullough and E.~Salvioni,
JHEP \textbf{12} (2022), 148
[erratum: JHEP \textbf{02} (2023), 165]
[arXiv:2209.00666 ].

\bibitem{ATLAS:2025eii}
G.~Aad \textit{et al.} [ATLAS and CMS],
[arXiv:2504.00672 [hep-ex]].


\bibitem{Anastasiou:2016cez}
C.~Anastasiou \textit{et al.},
\emph{JHEP} {\bf 05} (2016) 058.

\bibitem{Mistlberger:2018etf}
B.~Mistlberger, 
\emph{JHEP} {\bf 05} (2018) 028.

\bibitem{Aglietti:2004nj}
U.~Aglietti, R.~Bonciani, G.~Degrassi and A.~Vicini,
Phys. Lett. B \textbf{595} (2004), 432-441
[arXiv:hep-ph/0404071].

\bibitem{Actis:2008ug}
S.~Actis \textit{et al.}, 
\emph{Phys.\ Lett.\ B} {\bf 670} (2008) 12.

\bibitem{Anastasiou:2015ema}
C.~Anastasiou {\it et al.}, 
\emph{JHEP} {\bf 05} (2016) 058.

\bibitem{Harlander:2009mq}
R.~Harlander, H.~Mantler, S.~Marzani and K.~Ozeren, 
\emph{Eur.\ Phys.\ J.\ C} {\bf 66} (2010) 359.

\bibitem{Mistlberger:2018etf}
B.~Mistlberger, 
\emph{JHEP} {\bf 05} (2018) 028.

\bibitem{Anastasiou:2018adr}
C.~Anastasiou, V.~del Duca, E.~Furlan, B.~Mistlberger, F.~Moriello, A.~Schweitzer and C.~Specchia,
JHEP \textbf{03} (2019), 162
[arXiv:1811.11211].

\bibitem{Czakon:2024ywb}
M.~Czakon, F.~Eschment, M.~Niggetiedt, R.~Poncelet and T.~Schellenberger,
JHEP \textbf{10} (2024), 210
[arXiv:2407.12413].

\bibitem{Czakon:2023kqm}
M.~Czakon, F.~Eschment, M.~Niggetiedt, R.~Poncelet and T.~Schellenberger,
Phys. Rev. Lett. \textbf{132} (2024) no.21, 211902
[arXiv:2312.09896].

\bibitem{McGowan:2022nag}
J.~McGowan, T.~Cridge, L.~A.~Harland-Lang and R.~S.~Thorne,
Eur. Phys. J. C \textbf{83} (2023) no.3, 185
[erratum: Eur. Phys. J. C \textbf{83} (2023) no.4, 302]
[arXiv:2207.04739].

\bibitem{Cridge:2024icl}
T.~Cridge, L.~A.~Harland-Lang, J.~McGowan, R.~S.~Thorne, R.~D.~Ball, A.~Candido, S.~Carrazza, J.~Cruz-Martinez, L.~Del Debbio and S.~Forte, \textit{et al.}
J. Phys. G \textbf{52} (2025), 065002
[arXiv:2411.05373 ].

\bibitem{NNPDF:2024nan}
R.~D.~Ball \textit{et al.} [NNPDF],
Eur. Phys. J. C \textbf{84} (2024) no.7, 659
[arXiv:2402.18635].


\bibitem{Borowka:2016ehy}
S.~Borowka {\it et al.},
\emph{Phys.\ Rev.\ Lett.} {\bf 117} (2016) 012001.

\bibitem{Baglio:2018lrj}
J.~Baglio, F.~Campanario, S.~Glaus, M.~M{\"u}hlleitner, M.~Spira and J.~Streicher,
Eur. Phys. J. C \textbf{79} (2019) no.6, 459
[arXiv:1811.05692].

\bibitem{Dawson:1998py}
S.~Dawson, S.~Dittmaier and M.~Spira,
Phys. Rev. D \textbf{58} (1998), 115012
[arXiv:hep-ph/9805244].

\bibitem{Grigo:2013rya}
J.~Grigo, J.~Hoff, K.~Melnikov and M.~Steinhauser,
Nucl. Phys. B \textbf{875} (2013), 1-17
[arXiv:1305.7340].

\bibitem{Degrassi:2016vss}
G.~Degrassi, P.~P.~Giardino and R.~Gr{\"o}ber,
Eur. Phys. J. C \textbf{76} (2016) no.7, 411
[arXiv:1603.00385].

\bibitem{Grober:2017uho}
R.~Gr{\"o}ber, A.~Maier and T.~Rauh,
JHEP \textbf{03} (2018), 020
[arXiv:1709.07799].

\bibitem{Bonciani:2018omm}
R.~Bonciani, G.~Degrassi, P.~P.~Giardino and R.~Gr{\"o}ber,
Phys. Rev. Lett. \textbf{121} (2018) no.16, 162003
[arXiv:1806.11564].

\bibitem{Davies:2018qvx}
J.~Davies, G.~Mishima, M.~Steinhauser and D.~Wellmann,
JHEP \textbf{01} (2019), 176
[arXiv:1811.05489].

\bibitem{Bellafronte:2022jmo}
L.~Bellafronte, G.~Degrassi, P.~P.~Giardino, R.~Gr{\"o}ber and M.~Vitti,
JHEP \textbf{07} (2022), 069
[arXiv:2202.12157].

\bibitem{Davies:2023vmj}
J.~Davies, G.~Mishima, K.~Sch{\"o}nwald and M.~Steinhauser,
JHEP \textbf{06} (2023), 063
[arXiv:2302.01356].

\bibitem{Bagnaschi:2023rbx}
E.~Bagnaschi, G.~Degrassi and R.~Gr{\"o}ber,
Eur. Phys. J. C \textbf{83} (2023) no.11, 1054
[arXiv:2309.10525].

\bibitem{Baglio:2020wgt}
J.~Baglio, F.~Campanario, S.~Glaus, M.~M{\"u}hlleitner, J.~Ronca and M.~Spira,
Phys. Rev. D \textbf{103} (2021) no.5, 056002
[arXiv:2008.11626].

\bibitem{Jaskiewicz:2024xkd}
S.~Jaskiewicz, S.~Jones, R.~Szafron and Y.~Ulrich,
JHEP \textbf{09} (2025), 015
[arXiv:2501.00587].

\bibitem{Davies:2023obx}
J.~Davies, K.~Sch{\"o}nwald and M.~Steinhauser,
Phys. Lett. B \textbf{845} (2023), 138146
[arXiv:2307.04796].

\bibitem{Davies:2024znp}
J.~Davies, K.~Sch{\"o}nwald, M.~Steinhauser and M.~Vitti,
JHEP \textbf{08} (2024), 096
[arXiv:2405.20372].

\bibitem{Davies:2025ghl}
J.~Davies, K.~Sch{\"o}nwald and M.~Steinhauser,
JHEP \textbf{08} (2025), 192
[arXiv:2503.17449].

\bibitem{Buchalla:2013eza}
G.~Buchalla, O.~Cat{\'a} and C.~Krause,
Phys. Lett. B \textbf{731} (2014), 80-86
[arXiv:1312.5624].

\bibitem{Brivioinprep}
I.~Brivio, R.~Gr{\"o}ber and K.~Schmid,
[arXiv:2511.23410].


\bibitem{Heinrich:2020ckp}
G.~Heinrich, S.~P.~Jones, M.~Kerner and L.~Scyboz,
JHEP \textbf{10} (2020), 021
[arXiv:2006.16877].

\bibitem{Heinrich:2022idm}
G.~Heinrich, J.~Lang and L.~Scyboz,
JHEP \textbf{08} (2022), 079
[erratum: JHEP \textbf{10} (2023), 086]
[arXiv:2204.13045].

\bibitem{Maltoni:2024dpn}
F.~Maltoni, G.~Ventura and E.~Vryonidou,
JHEP \textbf{12} (2024), 183
[arXiv:2406.06670].

\bibitem{Heinrich:2024rtg}
G.~Heinrich and J.~Lang,
SciPost Phys. \textbf{18} (2025), 113
[arXiv:2409.19578].

\bibitem{DiNoi:2023ygk}
S.~Di Noi, R.~Gr{\"o}ber, G.~Heinrich, J.~Lang and M.~Vitti,
Phys. Rev. D \textbf{109} (2024) no.9, 095024
[arXiv:2310.18221].

\bibitem{DiNoi:2024ajj}
S.~Di Noi, R.~Gr{\"o}ber and M.~K.~Mandal,
JHEP \textbf{12} (2025), 220
[arXiv:2408.03252].

\bibitem{DiNoi:2025tka}
S.~Di Noi, B.~A.~Erdelyi and R.~Gr{\"o}ber,
[arXiv:2510.14680].

\bibitem{DiNoi:2025arz}
S.~Di Noi and R.~Gr{\"o}ber,
Phys. Lett. B \textbf{869} (2025), 139878
[arXiv:2507.10295].

\bibitem{Duhr:2025yor}
C.~Duhr, G.~Ventura and E.~Vryonidou,
JHEP \textbf{11} (2025), 046
[arXiv:2508.04500 ].

\bibitem{Chanowitz:1979zu}
M.~S.~Chanowitz, M.~Furman and I.~Hinchliffe,
Nucl. Phys. B \textbf{159} (1979), 225-243.

\bibitem{tHooft:1972tcz}
G.~'t Hooft and M.~J.~G.~Veltman,
Nucl. Phys. B \textbf{44} (1972), 189-213.
\bibitem{Breitenlohner:1977hr}
P.~Breitenlohner and D.~Maison,
Commun. Math. Phys. \textbf{52} (1977), 11-38.

\bibitem{DiNoi:2025uan}
S.~Di Noi, R.~Gr{\"o}ber and P.~Olgoso,
JHEP \textbf{09} (2025), 027
[arXiv:2504.00112].

\bibitem{Fuentes-Martin:2025meq}
J.~Fuentes-Mart{\'\i}n, A.~Moreno-S{\'a}nchez and A.~E.~Thomsen,
[arXiv:2507.19589].

\bibitem{DiNoi:2025uhu}
S.~Di Noi, H.~El Faham, R.~Gr{\"o}ber, M.~Vitti and E.~Vryonidou,
[arXiv:2507.01137].

\bibitem{Grober:2025vse}
R.~Gr{\"o}ber, A.~N.~Rossia and M.~Ryczkowski,
[arXiv:2509.02680].

\bibitem{Brivioinprep2}
I.~Brivio, R.~Gr{\"o}ber and K.~Schmid,
[arXiv:2511.23411].

\bibitem{Carvalho:2015ttv}
A.~Carvalho, M.~Dall'Osso, T.~Dorigo, F.~Goertz, C.~A.~Gottardo and M.~Tosi,
JHEP \textbf{04} (2016), 126
[arXiv:1507.02245].

\bibitem{Capozi:2019xsi}
M.~Capozi and G.~Heinrich,
JHEP \textbf{03} (2020), 091
[arXiv:1908.08923].

\bibitem{Alasfar:2023xpc}
L.~Alasfar, L.~Cadamuro, C.~Dimitriadi, A.~Ferrari, R.~Gr{\"o}ber, G.~Heinrich, T.~I.~Carlson, J.~Lang, S.~{\"O}rdek and L.~P.~S{\'a}nchez, \textit{et al.}
SciPost Phys. Comm. Rep. \textbf{2024} (2024), 2
[arXiv:2304.01968].



\bibitem{deBlas:2019rxi}
J.~de Blas, M.~Cepeda, J.~D'Hondt, R.~K.~Ellis, C.~Grojean, B.~Heinemann, F.~Maltoni, A.~Nisati, E.~Petit and R.~Rattazzi, \textit{et al.}
JHEP \textbf{01} (2020), 139
[arXiv:1905.03764 ].


\bibitem{Bishara:2016jga}
F.~Bishara, U.~Haisch, P.~Monni and E.~Re, 
\emph{Phys.\ Rev.\ Lett.} {\bf 118} (2017) 121801.

\bibitem{Soreq:2016rae}
Y.~Soreq, H.~X.~Zhu and J.~Zupan, 
\emph{JHEP} {\bf 12} (2016) 045.

\bibitem{Yu:2016rvv}
F.~Yu,
JHEP \textbf{02} (2017), 083
[arXiv:1609.06592].

\bibitem{Aguilar-Saavedra:2020rgo}
J.~A.~Aguilar-Saavedra, J.~M.~Cano and J.~M.~No,
Phys. Rev. D \textbf{103} (2021) no.9, 095023
[arXiv:2008.12538].

\bibitem{Falkowski:2020abc}
A.~Falkowski {\it et al.}, 
\emph{JHEP} {\bf 04} (2021) 020.

\bibitem{Alasfar:2019pmn}
L.~Alasfar, R.~Corral Lopez and R.~Gr{\"o}ber,
JHEP \textbf{11} (2019), 088
[arXiv:1909.05279].

\bibitem{Alasfar:2022vqw}
L.~Alasfar, R.~Gr{\"o}ber, C.~Grojean, A.~Paul and Z.~Qian,
JHEP \textbf{11} (2022), 045
[arXiv:2207.04157].

\bibitem{Balzani:2023jas}
E.~Balzani, R.~Gr{\"o}ber and M.~Vitti,
JHEP \textbf{10} (2023), 027
[arXiv:2304.09772].

\bibitem{Egana-Ugrinovic:2019dqu}
D.~Egana-Ugrinovic, S.~Homiller and P.~R.~Meade,
Phys. Rev. D \textbf{100} (2019) no.11, 115041
[arXiv:1908.11376].

\bibitem{Bar-Shalom:2018rjs}
S.~Bar-Shalom and A.~Soni,
Phys. Rev. D \textbf{98} (2018) no.5, 055001
[arXiv:1804.02400].

\bibitem{Nir:2024oor}
Y.~Nir and P.~P.~Udhayashankar,
JHEP \textbf{06} (2024), 049
[arXiv:2404.16545].

\bibitem{Giannakopoulou:2024unn}
A.~S.~Giannakopoulou, P.~Meade and M.~Valli,
JHEP \textbf{02} (2025), 067
[arXiv:2410.05236].

\bibitem{Erdelyi:2024sls}
B.~A.~Erdelyi, R.~Gr{\"o}ber and N.~Selimovic,
JHEP \textbf{05} (2025), 189
[arXiv:2410.08272].

\bibitem{deBlas:2017xtg}
J.~de Blas, J.~C.~Criado, M.~Perez-Victoria and J.~Santiago,
JHEP \textbf{03} (2018), 109
[arXiv:1711.10391].

\bibitem{CMS:2025xkn}
V.~Chekhovsky \textit{et al.} [CMS],
[arXiv:2502.05665].


\bibitem{Davoudiasl:2023huk}
H.~Davoudiasl and P.~P.~Giardino,
Phys. Rev. D \textbf{109} (2024) no.7, 075037
[arXiv:2311.12112].

\bibitem{Erdelyi:2025axy}
B.~A.~Erdelyi, R.~Gr{\"o}ber and N.~Selimovic,
JHEP \textbf{05} (2025), 135
[arXiv:2501.07628].

\bibitem{Allwicher:2025mmc}
L.~Allwicher, M.~McCullough, S.~Renner, D.~Rocha and B.~Smith,
[arXiv:2511.02642].


\bibitem{dEnterria:2021abc}
D.~d Enterria, J.~Poldaru and K.~Wojcik, 
\emph{JHEP} {\bf 10} (2021) 188.

\end{thebibliography}
\end{document}